\documentclass[pra,reprint,superscriptaddress,showpacs,amsmath, floatfix,amssymb,aps]{revtex4-1}
\usepackage{graphicx}% Include figure files
\usepackage{subfigure}
\usepackage{dcolumn}% Align table columns on decimal point
\usepackage{bm}% bold math
\usepackage{hyperref}% add hypertext capabilities
\usepackage{indentfirst}
\usepackage{braket}
\usepackage{amsmath}
\usepackage{ulem}
\usepackage{color}
\hypersetup{colorlinks=true, citecolor=blue, urlcolor=blue, linkcolor=blue}
\begin{document}

\title{Enhancement of boson superfluidity in a one-dimensional Bose-Fermi mixture}
\author{Chenrong Liu}
\affiliation{Department of Physics and State Key Laboratory of Surface Physics, Fudan University, Shanghai 200433, China}
\author{Yongzheng Wu}
\affiliation{The 32nd Research Institute of China Electronics Technology Group Corporation, Shanghai 200433, China}
\author{Jie Lou}
\email{loujie@fudan.edu.cn}
\affiliation{Department of Physics and State Key Laboratory of Surface Physics, Fudan University, Shanghai 200433, China}
\author{Yan Chen}
\email{yanchen99@fudan.edu.cn}
\affiliation{Department of Physics and State Key Laboratory of Surface Physics, Fudan University, Shanghai 200433, China}

\date{\today}
\begin{abstract}
We examine the effect of boson-fermion interaction in a one-dimensional Bose-Fermi mixture by using the density matrix renormalization group method. We show that the boson superfluidity is enhanced by fermions for a weak boson-fermion coupling at an approximate integer boson filling factor (e.g., $0.935\le \rho_b \le 1.0$), and this enhancement is produced both in a fermion metallic state and in a fermion insulating state.  A metal-insulator phase transition of fermions induced by boson-fermion interaction is observed even though there is no fermion-fermion interaction in the parent Hamiltonian. Furthermore, we find that the boson superfluid order and density wave order can coexist in a deep fermion Mott region.  All these features could be measured in future experiments and open up the possibility of detecting the new physical effect in the Bose-Fermi mixture.
\end{abstract}
\pacs{}
% insert suggested keywords - APS authors do not need to do this
%\keywords{}
\maketitle

\section{Introduction}
Bose-Fermi mixture, which has been recently realized in ultracold atoms experiments \cite{Modugno2240, PhysRevLett.92.140405, Ferrier-Barbut1035}, represents a new system for studying strongly correlated many-body physics.  The physics of the one-dimensional Bose-Fermi mixture was extensively studied based on theories beyond the mean-field approximation or the perturbation theory \cite{PhysRevA.73.021602, PhysRevA.77.012115, PhysRevA.77.023601,  PhysRevA.77.041608, PhysRevA.79.053604, PhysRevLett.96.190402, PhysRevA.72.061603, PhysRevLett.93.120404,PhysRevA.75.013612,PhysRevLett.91.150403}. In such a mixture, the effective interaction induced by the boson-fermion interaction plays an essential role in the emergence of a new phase and new physical effect. For the effect of bosons on fermions, Kinnunen \textit{et al.} \cite{PhysRevLett.121.253402}  presented a strong coupling theory for the critical temperature of  $p$-wave pairing between spin-polarized (spinless) fermions immersed in a Bose-Einstein condensate.  Quite recently, a quantum Monte Carlo study on a two-dimensional Bose-Fermi mixture revealed that an effective $p$-wave interaction between fermions will be induced as far as the bosons are in a superfluid state, and the composite fermion pairs may appear at low temperatures \cite{pwave}. For the effect of fermions on bosons, an investigation of the effective interaction between bosons was previously carried out in a two-dimensional Bose-Fermi mixture by using the linear response theory \cite{PhysRevLett.91.130404}. Their results show that the second-order term of the effective boson interaction can be attractive.  This attractive boson effective interaction was also investigated using phenomenological bosonization and Green's function techniques in one dimension \cite{PhysRevA.81.053626}.  Numerically, a quantum Monte Carlo study of this induced boson-boson interaction was proposed in the normal fermi state for a narrow boson-fermion coupling region \cite{PhysRevA.77.023608}, and they show that the boson superfluid phase region expands. Moreover, the fermion-mediated interactions between bosonic atoms have been observed in experiments. Their results indicate that when a Bose-Einstein condensate of Cesium atoms is embedded in a degenerate Fermi gas of lithium atoms, interspecies interactions can give rise to an attractive boson-boson interaction \cite{DeSalvo}. However,  few theoretical works have been done on how the particle filling factor and fermion state influence the effective boson-boson interaction.   \\

It is well known that in the one-dimensional Bose-Hubbard model, there exists a quantum phase transition from superfluid phase to the Mott insulator phase when the onsite boson repulsive interaction $U_{ {bb}}$ switches on \cite{PhysRevA.85.053644, PhysRevB.61.12474}. On the other hand, there is a  metal-insulator phase transition in the one-dimensional spinless Fermi-Hubbard model with a repulsive nearest-neighbor interaction $V_{ {ff}}$ at half fermion filling \cite{PhysRevLett.49.1691}. In the weak-coupling limit $U_{ {bf}} \rightarrow 0$,  one can expect that bosons and fermions are decoupled, e.g., the ground state of the mixture is a combination of boson superfluid and fermi metal. Otherwise,  the mixture is in a phase separation state for the strong coupling limit $U_{ {bf}} \rightarrow \infty$ (see Appendix). The effective boson-boson (fermion-fermion) interaction, induced by the exchange of fermions (bosons), leads to new phases and new effects between these two limits.  \\

In this paper, we mainly study the physical effect induced by the effective boson-boson interaction in a one-dimensional Bose-Fermi mixture. We find that the boson superfluidity is enhanced at a weak repulsive onsite boson-fermion coupling (e.g. $U_{ {bf}}/t < 3.0$) when the boson filling factor $\rho_b$ is 1.0 or extremely close to this integer ($0.935\le \rho_b \le 1.0$), here $t$ is the boson-boson and fermion-fermion nearest-neighbor hopping amplitude. The correlation functions and fermi momentum distribution are calculated at different filling factors using the density matrix renormalization group (DMRG) method. To analyze how different fermion states affect the enhancement of boson superfluidity, the interacting and non-interacting fermions are both considered in our model. In the strong boson-fermion coupling limit, the boson density distribution in real space is also addressed.

\section{Model and method}
We consider a mixture of spinless fermions and bosons in one dimension. The canonical ensemble Hamiltonian reads
\begin{eqnarray} \label{Ham}
H =&&H_t+H_{\textup{int}} , \nonumber \\
H_{t} =&&-\sum_{\langle i j\rangle}\left(t_ {b} b_{i}^{\dagger} b_{j}+t_{f} c_{i}^{\dagger} c_{j}+\textup{H. c.}\right) ,  \\
H_{\textup{int}} =&&\frac{U_{ {bb}}}{2} \sum_{i} n_{i}^{ {b}}\left(n_{i}^{ {b}}-1\right)+U_{ {bf}} \sum_{i} n_{i}^{ {b}} n_{i}^{ {f}}   \nonumber \\
              &&+ V_{ {ff}} \sum_{\langle i j\rangle} n_{i}^{ {f}} n_{j}^{ {f}} .  \nonumber
\end{eqnarray}
where $t_{b}$($t_{f}$) is the boson-boson (fermion-fermion) nearest-neighbor hoping amplitude, $b_i$($c_i$) is the boson (fermion) annihilation operator, $U_{{bb}}$($U_{{bf}}$) is the onsite boson-boson (boson-fermion) repulsive interaction, $V_{{ff}}$ is Coulomb repulsive interaction between the nearest-neighbor fermions,   and $n_{i}^{{b}}$($n_{i}^{{f}}$) is the onsite boson (fermion) number operator. \\

In our calculations, $t_{b}=t_{f}=t$ is set to 0.1. To study the effect of fermions on bosons, we fixed $U_{{bb}}/t=3.0$ which is close to a Mott phase transition point of the one dimensional Bose-Hubbard model.  In the decoupled case, bosons form a superfluid state at filling factor $\rho_b=1$ because $U_{{bb}}/t=3.0$ is lower than the superfluid-Mott transition point $U_{{bb}}^{\textup{c}}/t\approx3.3$ \cite{PhysRevB.71.104508}. And fermions would constitute either a metallic state with $V_{{ff}}/t < 2.0$ or a Mott insulating state with $V_{{ff}}/t > 2.0$ at fermion half filling\cite{Des}.  In the following, we study two distinct cases: free fermions ($V_{{ff}}/t=0$) and interacting fermions ($V_{{ff}}/t>0$). The boson (fermion) filling factor is defined as $\rho_{{b(f)}}=N_{{b(f)}}/N_{{s}}$.  Besides, the lattice size $N_{s}$ is set as 62 for fermion half filling $\rho_{f}=1/2$ and $N_{s}=60$ for fermion quarter filling $\rho_{f}=1/4$. The reason why we set the lattice size in this way is that we should keep the fermion particle number odd to remove fermion degeneracy in the ground state.  For bosons, we set an occupy number cutoff $n_{{b}}^{\textup{cut}}=2$  per site on account of the boson repulsive interaction $U_{{bb}}$.\\

We use the DMRG method \cite{PhysRevB.48.10345}  to perform the numerical calculations.  The maximum number of kept states is $4500$, and we use up to $40$ sweeps. Besides, we add some noises in the first nine sweeps to avoid a metastable state. Under the setup, the largest truncation error during the sweep is $\sim 10^{-7}$, and the energy convergence precision is $\sim 10^{-8}$.  All the calculations were conducted using the ITensor library \cite{ITensor} with periodic boundary condition.

\section{Numerical results}
In this part, we will discuss our numerical results in two subsections. One is for free fermions with $V_{{ff}}=0$, and the other is for interacting fermions with $V_{{ff}}>0$.   In the free fermion case, we mainly study the fermion-mediated effective boson-boson interaction and investigate the effect of boson-fermion coupling $U_{{bf}}$ at different particle filling factors. We focus on the relations between the boson-boson correlation function and the fermion state in the interacting fermion section.  \\

\subsection{Free fermions: $V_{{ff}}/t=0$}
%%%%% Figure1 %%%%%%%
\begin{figure*}[htb]
\centering
\includegraphics[width=0.9\textwidth]{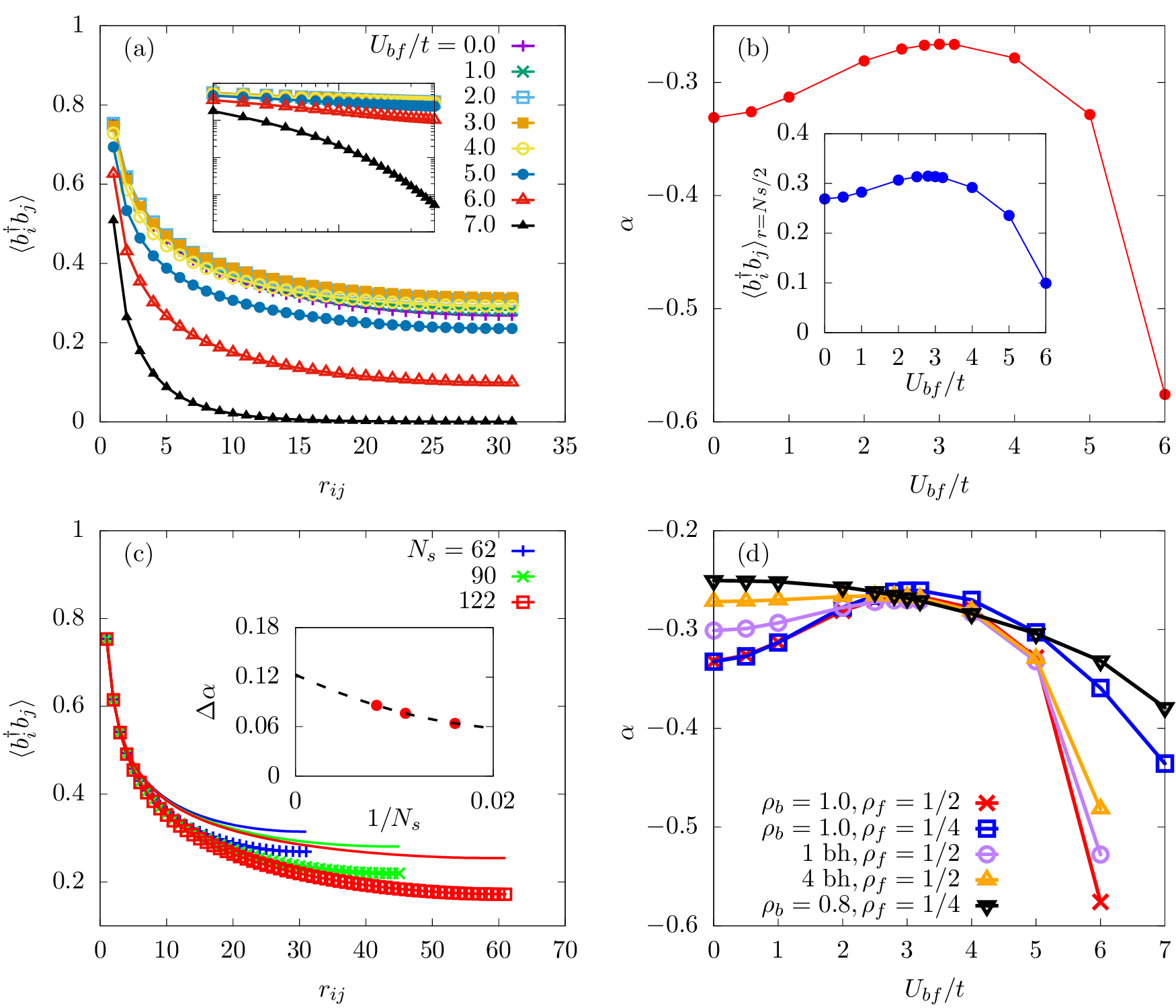}
\caption{\label{Fig1}(Color online) Boson-boson correlation functions and its decay exponents. (a)-(c) are calculated with fixed particle filling factors, e.g. $\rho_{ {b}}=1.0$ and $\rho_{ {f}}=1/2$. (a) The boson-boson correlation functions $\langle b_i^\dag b_j\rangle$ vs lattice distance $r_{ij}$ for different values of $U_{{bf}}/t$, the insert figure is plotted by using a logarithmic scale for both the $x$-axis and the $y$-axis. (b) The power law decay exponents $\alpha$ of (a), the insert figure is the value of  $\langle b_i^\dag b_j \rangle$ at the distance $r=N_{{s}}/2$. (c) Finite size effect of boson-boson correlation function. The full line and the line-symbol indicates the boson-fermion coupling is 2.8 and 0 respectively. The insert figure shows the difference of $\alpha$ between $U_{{bf}}/t=2.8$ and $U_{{bf}}/t=0$, the dash line is a fitting function. (d) $\alpha$ as a function of $U_{{bf}}/t$ at different boson and fermion filling factors. Here, 1bh represents $\rho_b=N_b/N_s=(N_{s}-1)/N_s=0.984$ and 4bh labels $\rho_b=N_b/N_s=(N_{s}-4)/N_s=0.935$.}
\end{figure*}
%%%%% Figure1 %%%%%%%
In order to study the influence of the onsite boson-fermion repulsive interaction. One of the primary efforts here is to obtain the boson-boson correlation function that corresponds to the boson superfluid order.  The correlations $\langle b_i^\dag b_j\rangle$ as a function of lattice distance $r_{ij}$ were calculated for several different values of $U_{ {bf}}/t$ at boson filling factor $\rho_{{b}}=1$ and fermion filling factor $\rho_{{f}}=1/2$ as shown in Fig.~\ref{Fig1}(a).  The curves here show that for $U_{{bf}}/t \leqslant 6.0 $,  a power-law decay behavior of $\langle b_i^\dag b_j\rangle$ is observed, which indicates that the boson is in a superfluid state. The power-law relation also reveals the so called quasi-long-range order which means there is no truly boson superfluidity in one dimension.  When $U_{{bf}}\gg t $, we expect that the superfluid should be broken by the fermions so that  $\langle b_i^\dag b_j\rangle$ decay exponentially. In our simulations, a density-wave state is established at $U_{{bf}}/t=7.0$ (see Appendix).     \\

From Fig.~\ref{Fig1}(a)-(b), we find that the boson superfluidity is enhanced when $U_{bf}/t$ is increased from 0 to 3.0.  It appears that the boson-boson correlation $\langle b_i^\dag b_j\rangle$ at distance $r=N_{{s}}/2$ in Fig.~\ref{Fig1}(a) rises as $U_{{bf}}/t$ is increased from $0$ to $3.0$,  and then this value decreases when $U_{{bf}}/t$ continue to grow up.  In other words, the boson-boson correlation function decays more slowly if $U_{{bf}}/t$ goes from $0$ to $3.0$.  To analyze the behavior of the decay rate quantitatively, we fit $\langle b_i^\dag b_j\rangle$ with a power-law relation,
\begin{eqnarray} \label{power_law}
\langle b_i^\dag b_j\rangle = A r_{ij}^\alpha,
\end{eqnarray}
where $\alpha$ is the decay exponent, and $A$ denotes a constant. Thus, $\alpha$ can be used to describe the strength of the boson superfluid order.  These exponents are shown in Fig.~\ref{Fig1}(b).  As one can see, $\alpha$ indeed increases in region $0<U_{{bf}}/t<3.0$ and it has a maximum value around $U_{{bf}}/t=3.0$.  \\

This enhancement can be understood by the perturbation theory \cite{PhysRevLett.91.130404,PhysRevA.69.063603,PhysRevB.80.054511}.  Within the theory, the boson drives the fermionic system, and a perturbed fermionic density, in turn, acts as a driver for the bosons. As a consequence of this feedback, an attractive interaction between bosons is induced, which can compensate the direct boson-boson repulsive interaction $U_{{bb}}$ and favor superfluidity. A simple picture enables this to be understood as follows: for the repulsive boson-fermion coupling,  fermions are repelled from the bosons, and thus the bosons feel an attractive interaction where the fermion density is quite low \cite{DeSalvo}. Therefore, this induced attractive boson-boson interaction would be determined by the boson-fermion interaction.\\

To check whether the enhancement is due to the finite size effect, we calculate the boson-boson correlation function for three different lattice sizes at a fixed boson and fermion filling factor. The results are shown in Fig.~\ref{Fig1}(c). Here, the separation of the two correlation functions is expanded for a larger lattice size, and the insert figure also indicates that the difference $\Delta \alpha$ is increased with a growing lattice size. The finite-size scaling of $\Delta \alpha$ which is presented in the insert figure of Fig.~\ref{Fig1}(c) predicts $\Delta \alpha =0.12$ when $Ns\rightarrow \infty$. This result clearly shows that the observed enhancement of the boson superfluidity does not come from the finite size effect. \\

If there are only a few bosons in the mixture, then the boson can not drive the fermionic system.  The total number of bosons would have an important role in the formation of the enhancement.  We examine $\alpha$ against $U_{{bf}}/t$ at different boson and fermion filling factors. We start with an integer boson filling factor $\rho_{{b}}=1$ and then gradually reduce the total boson number.  Numerical results are presented in Fig.~\ref{Fig1}(d). One can observe that $\alpha$ is increased in a weak bose-fermi coupling region when the boson filling factor is an integer($\rho_{{b}}=1$) or extremely close to this integer.  For boson filling factor $\rho_b=1$, this increment behavior exists at both a half and a quarter fermion filling factor. But for a non-integer boson filling factor, e.g. $\rho_{{b}}=0.8$, $\alpha$ decreases monotonically in the whole range of $U_{{bf}}/t$. These features show that the boson superfluidity enhancement can only exist at $\rho_b=1.0$ or a fractional boson filling factor but very close to 1.0. The reason why the enhancement dependent on the boson filling factor is the following. There are two competing orders for an integer boson filling factor, namely superfluid and Mott insulator, in the mixture. In contrast, the Mott insulator state is such a state that the boson number per site is an integer, and there is no boson number fluctuation among the sites. When $U_{bf}$ is added, the bosons would feel a repulsive interaction on the fermion occupied site, and thus the boson number fluctuation among the sites is induced. In other words, the Mott phase region is shrunk, and thus the enhancement is observed. However, this is only true for a boson filling factor close to an integer. At a fractional boson filling factor($\rho_b=0.8$), there is no Mott insulator state to start with, and the scattering of the fermions should always suppress the boson superfluidity.     \\

%%%%%Figure 2%%%%
\begin{figure}[htb]
\centering
\includegraphics[width=0.45\textwidth]{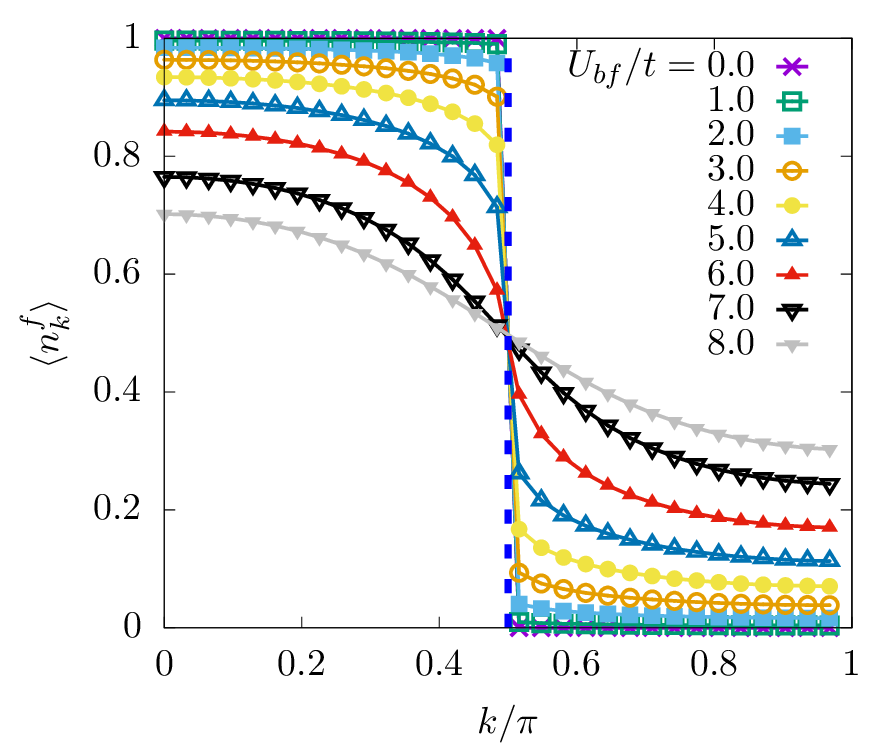}
\caption{\label{Fig2}(Color online) The fermion particle distribution in momentum space at a fixed boson filling factor $\rho_{ {b}}=1.0$ and fermion filling factor $\rho_{ {f}}=1/2$. The blue dashed vertical line is the position of Fermi momentum $k_F/\pi=1/2$. }
\end{figure}
%%%%%%%%%%%%%
The effect of bosons on fermions is also addressed. We calculate the fermion momentum distribution $\langle n_k^{{f}} \rangle$ in Fig.~\ref{Fig2} at a fixed boson filling factor $\rho_{{b}}=1.0$ and fermion filling factor $\rho_{{f}}=1/2$.  The $\langle n_k^{{f}}\rangle$ is defined as,
\begin{eqnarray} \label{nk}
\langle n_k^{ {f}} \rangle= \frac{1}{N_{ {s}}}\displaystyle{\sum_{i,j}} \textup{e}^{-i\vec{k}\cdot\left(\vec{r}_i-\vec{r}_j\right)}\langle c_i^\dag c_j\rangle,
\end{eqnarray}
which is the Fourier transformation of the equal time Green's function \cite{QPOD}, and gives the occupation of fermions in momentum space. \\

As the violet line-cross of Fig.~\ref{Fig2} shows, all the fermions are free when $U_{{bf}}/t$ is zero  and this leads to the fact that $\langle n_k^{{f}} \rangle$ has a discontinuity ($Z=1$) at the corresponding Fermi momentum $k_F/\pi=1/2$.  When $U_{ {bf}}/t$ turns on and rises up, instead of a jump, one could finds a power-law singularity at $k=k_F$.  This is the feature of a Tomonaga-Luttinger liquid (TLL) \cite{LL}, and it means that there exists an effective interaction between fermions, which is induced by the exchange of bosons. If the boson-fermion coupling is too strong, e.g., $U_{{bf}}/t \geqslant 7.0$, then the fermions and bosons can not occupy the same site, and the fermions exhibit a charge density wave (CDW) pattern in real space (See Fig.~\ref{FigS2}(b) in Appendix). Consequently, the singularity does not exist as the empty black triangle-line of Fig.~\ref{Fig2} shown. This evidence shows that although there is no fermion-fermion interaction in the Hamiltonian, there is still an induced phase transition from a TLL phase to a CDW phase when we switch on $U_{{bf}}$.

\subsection{Interacting fermions: $V_{ {ff}}/t > 0$}
\begin{figure}[htb]
\centering
\includegraphics[width=0.45\textwidth]{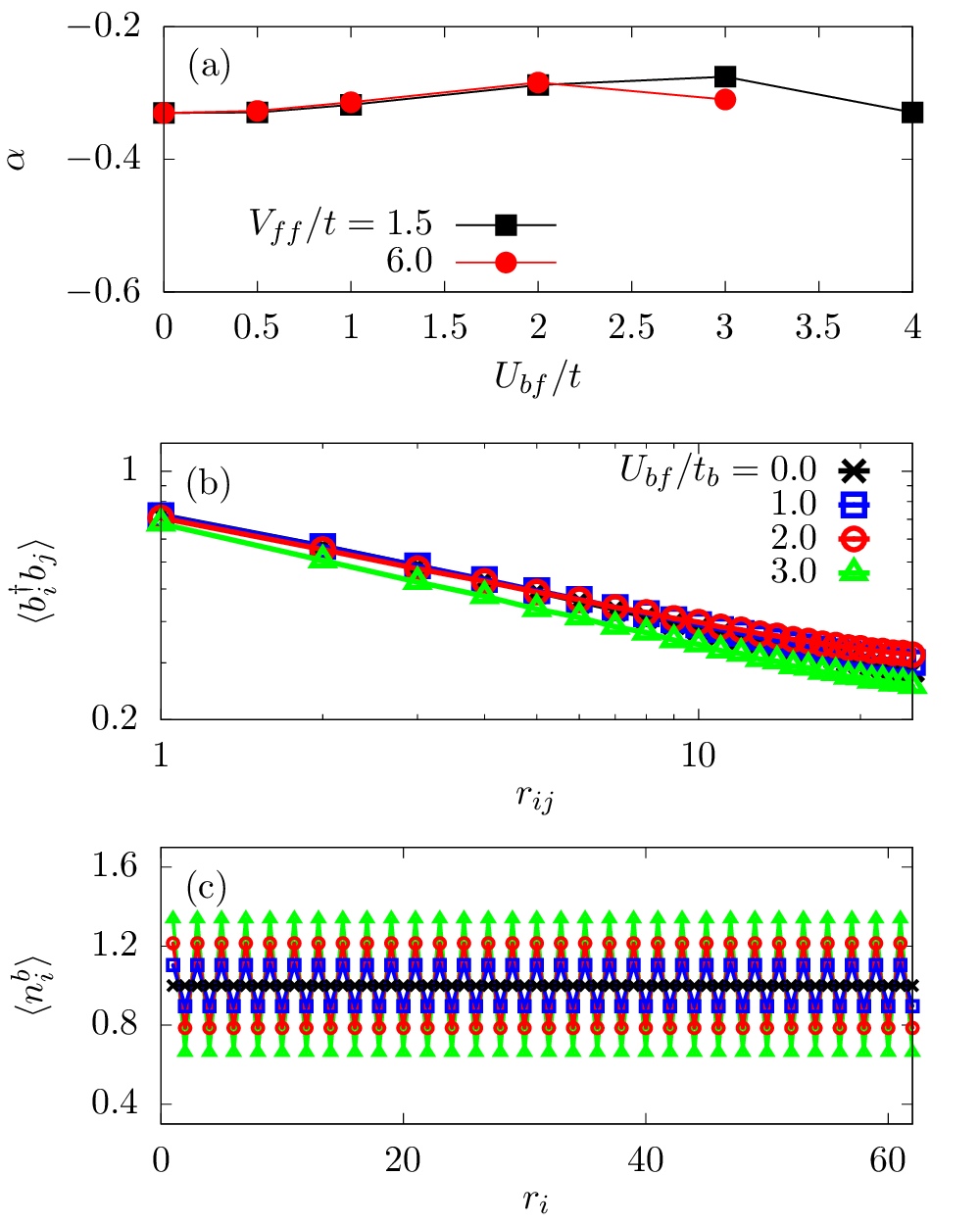}
\caption{\label{Fig3}(Color online) The exponents, correlation functions and particle density distribution are calculated at finite $V_{ {ff}}$. (a) $\alpha$ as a function of $U_{ {bf}}/t$ for fermions in a metallic state (black line) and an insulator state (red line). (b) Boson-boson correlation functions and (c) boson density distribution in real space are evaluated in a deep fermion insulator region where the fermion hopping $t_f$ is set to be zero and $V_{ff}/t_b=100$. The same line color and symbol in (b)-(c) represents the same value of $U_{ {bf}}$. }
\end{figure}

In this section, we discuss the effect of interacting fermions on bosons by setting the fermion interaction $V_{ {ff}}$ to be non-zero. The particle filling factors here are fixed at $\rho_{ {b}}=1$ and $\rho_{ {f}}=1/2$.  Under these conditions, we can tune the fermion state from a metal to an insulator and examine how different fermion states affect the boson superfluidity. The result is presented in Fig.~\ref{Fig3}(a).  Interestingly,  the enhancement of boson superfluidity is observed in both situations of the fermion state.  For the fermion insulator case, the scattering of fermions still can expand the boson superfluid phase region, and thus the boson superfluidity is enhanced. However, for strong boson-fermion coupling, bosons are localized. \\
%%%%%Figure 3%%%%
%%%%%%%%%%%%%

If fermions are in a deep insulating region, fermions are pinned on the lattice site. The bosons should have a density wave feature in real space at a finite $U_{ {bf}}$.  Therefore, the boson superfluid order and density wave order can coexist. We examine this by calculating the boson correlation function and boson density distribution. Here, we set fermion-fermion hopping $t_{ {f}}=0$ and fermion-fermion nearest-neighbor repulsive interaction $V_{ {ff}}/t_{ {b}}=100$.  The results are shown in Fig.~\ref{Fig3}(b) and (c).  For a weak boson-fermion coupling $U_{ {bf}}$ (e.g. $U_{ {bf}}/t_{ {b}}=1.0$)  a boson-boson correlation function with power-law decay is observed in Fig.~\ref{Fig3}(b) indicating that the bosons are in a superfluid state. On the other hand, its density distribution in real space (blue curve in Fig.~\ref{Fig3}(c)) clearly shows a density wave feature. These results indicate that the boson superfluid order coexists with the density wave order in the deep fermion insulator region.   \\

\section{Conclusions}
We have studied the enhancement of boson superfluidity in a one-dimensional Bose-Fermi mixture for free and interacting fermions at different particle filling factors.  Close to an integer boson filling factor (e.g., $0.935\le \rho_b \le 1.0$), our results show that the boson superfluidity is enhanced for weak boson-fermion coupling. Meanwhile, this enhancement effect is observed in a fermion metallic state and a fermion insulating state.  On the other hand, we show that the effective fermion-fermion interaction is induced at finite $U_{ {bf}}$ although $V_{ {ff}}$ is zero and thus leading to a phase transition from a fermion metallic state to a fermion CDW state.  The boson-boson correlation function and its density distribution are also calculated for interacting fermions in a very deep fermion insulator region with $t_{ {f}}=0 $ and $V_{ {ff}}/t_{ {b}}=100$, and we find that the boson superfluid order and density wave order coexist there. Recently, the boson-fermion interactions can be tuned in $^{87}$Rb - $^{40}$K mixtures via Feshbach resonance technology \cite{PhysRevLett.102.030408}. Our predictions could be tested in future experiments.

\section{Acknowledgments}
We are thankful for the useful discussions with T. K. Lee, X. W. Guan, and C. S. Ting. This work is supported by the National Key Research and Development Program of China (Grants Nos. 2017YFA0304204 and 2016YFA0300504), the National Natural Science Foundation of China Grant No. 11625416, and the Shanghai Municipal Government (Grants Nos. 19XD1400700 and 19JC1412702).

\appendix
\section{Comparing with quantum Monte Carlo method(QMC).}
%%%%%Figure 4%%%%
\begin{figure}[htb]
\centering
\includegraphics[width=0.45\textwidth]{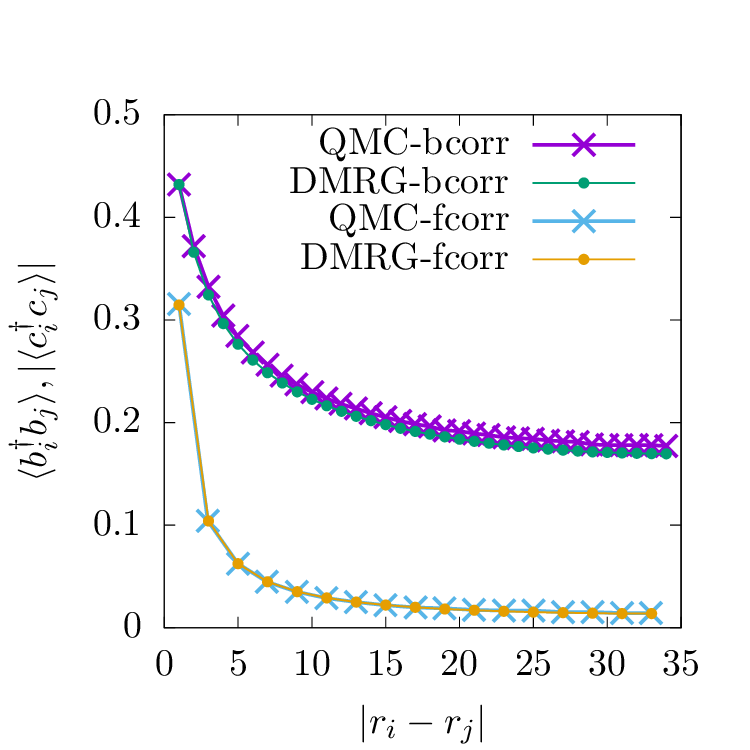}
\caption{\label{FigS1}(Color online) Comparing with quantum Monte Carlo method(QMC). The bcorr (fcorr) represents boson-boson (fermion-fermion) correlation function. QMC data is from Ref.\cite{PhysRevLett.96.190402}. }
\end{figure}
%%%%%%%%%%%%%
In order to compare with QMC, the parameters in Hamiltonian Eq.(\ref{Ham}) are set as $t_{ {b}}=t_{ {f}}=1.0$, $N_{ {s}}=70$, $U_{ {bb}}=5.0$, $V_{ {ff}}=0$ and $U_{ {bf}}=1.0$. The particle filling factors are fixed at $\rho_{ {f}}=\rho_{ {b}}=N_{ {s}}/2$.   These results are shown in Fig.~\ref{FigS1}. Because of the fractional $1/2$ boson filling, the bosons should form a superfluid state and thus the boson correlation function has a power-law decay behavior. For fermions, they are in a TLL phase.  All the DMRG calculations are consistent with that of QMC. \\
\section{Phase separation and induced fermion density wave state in the free fermion case. }
%%%%%Figure 5%%%%
\begin{figure}[htb]
\centering
\includegraphics[width=0.45\textwidth]{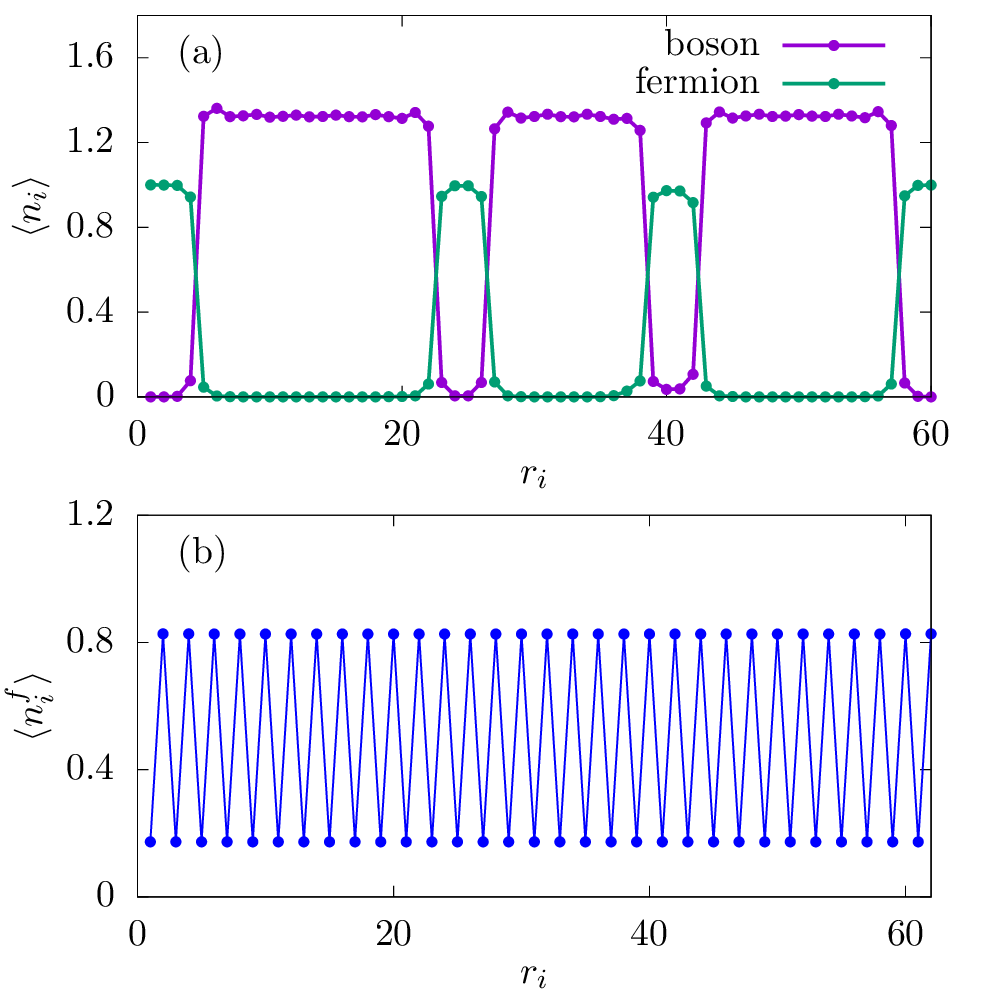}
\caption{\label{FigS2}(Color online) (a) Boson density distribution in real space with $U_{ {bf}}/t=10$, $\rho_{ {b}}=1.0$, $\rho_{ {f}}=1/4$ and (b) fermion density distribution in real space with $U_{ {bf}}/t=7.0$, $\rho_{ {b}}=1.0$, $\rho_{ {f}}=1/2$.}
\end{figure}
%%%%%%%%%%%%%
In this part, we calculate the particle distribution in real space at a strong boson-fermion coupling with fixed $V_{ff}/t=0$. The results are presented in Fig.~\ref{FigS2}.  When particle filling factors are set to be $\rho_{ {b}}=1.0$ and $\rho_{ {f}}=1/4$, the bosons and fermions are localized and in a phase separation state at $U_{bf}/t=10$ as shown in Fig.~\ref{FigS2}(a).  As one can see, the particle density in real space shows a cluster-like distribution, and the boson clusters are incompatible with fermion clusters. If we set $\rho_{ {b}}=1.0$ and $\rho_{ {f}}=1/2$, then a fermion CDW state is observed at $U_{bf}/t=7$ as produced in Fig.~\ref{FigS2}(b). All these features indicate that fermions can still obtain an effective interaction via the exchange of bosons.

\end{document}